\documentstyle[12pt,aasms4]{article}

\begin{document}
\newcommand{\newc}{\newcommand}
 
\newc{\be}{\begin{equation}}
\newc{\ee}{\end{equation}}
\newc{\ba}{\begin{eqnarray}}
\newc{\ea}{\end{eqnarray}}
\newc{\ie}{{\it i.e.}}
\newc{\eg}{{\it eg.}}
\newc{\etc}{{\it etc.}}
\newc{\etal}{{\it et al.}}
 
\newc{\ra}{\rightarrow}
\newc{\lra}{\leftrightarrow}
\newc{\no}{Nielsen-Olesen }
\newc{\lsim}{\buildrel{<}\over{\sim}}
\newc{\gsim}{\buildrel{>}\over{\sim}}
%AAS\TeX\
\begin{flushright}
Crete - 97/14
\end{flushright}
\title{A Statistic for the Detection of Long Strings in Microwave 
Background Maps.
}
\author{Leandros Perivolaropoulos}
\affil{Department of Physics,
University of Crete,
71003 Heraclion,
GREECE}
\authoremail{leandros@physics.uch.gr}

\begin{abstract}
Using analytical methods and Monte Carlo simulations, we analyze a new statistic 
designed to detect isolated step-like discontinuities which are coherent over 
large areas of Cosmic Microwave Background (CMB) pixel maps.
Such coherent temperature discontinuities are predicted by the {\it 
Kaiser-Stebbins} effect to form due to long cosmic strings present in our 
present horizon. The background of the coherent step-like seed is assumed to be 
a scale invariant Gaussian random field which could have been produced by a 
superposition of seeds on smaller scales and/or by inflationary quantum 
fluctuations. 
The effects of uncorrelated Gaussian random noise are also considered.
The statistical variable considered is the Sample Mean Difference (SMD) between 
large neighbouring sectors of CMB maps, separated by a straight line in two 
dimensional maps and a point in one dimensional maps.
We find that including noise, the SMD statistics can 
detect {\it at the $1\sigma$ to $2\sigma$ 
level} the 
presense of a long string with $G\mu (v_s \gamma_s)={1 \over {8\pi}}({{\delta 
T}\over T})_{rms} \simeq 0.5 \times 10^{-7}$ while more conventional statistics 
like the skewness or the kurtosis require a value of $G\mu$ {\it almost an order of 
magnitude larger} for detectability at a comparable level.  
\end{abstract}


\section{\bf Introduction}

The major progress achieved during the past 15 years in both theory and  
cosmological observations has turned the search for the origin of
cosmic structure into one of the most exciting fields of scientific 
research \markcite{e89}(for a good review see Efstathiou 1989). 
Despite the severe constraints imposed by detailed observational
data on theories for structure formation the central question remains open:
{\it What is
the origin of primordial fluctuations that gave rise to structure in the 
universe?} 
 Two classes of theories attempting to answer this question have emerged during 
the past ten years
and have managed to survive through the observational constraints with only 
minor adjustments.
 
According to the first class, primordial fluctuations are produced by quantum 
fluctuations of a
linearly coupled scalar field during a period of inflation \markcite{h82,s82,gp82,bs83}(Hawking 1982; 
Starobinsky 1982; Guth \&
Pi 1982; Bardeen, Steinhardt \& Turner 1983). These
fluctuations are subsequently expected to become classical and provide the 
progenitors of structure
in the universe. Because of the extremely small linear coupling of the scalar 
field, needed to
preserve the observed large scale homogeneity, the inflationary perturbations 
are expected by the
central limit theorem, to obey Gaussian statistics. This is not the case for the 
second class of
theories.
 
According to the second class of theories \markcite{k76,v81,v85,t89,b92,p94}(Kibble 1976; Vilenkin 1981; Vilenkin 
1985; Turok 1989; 
Brandenberger 1992; Perivolaropoulos 1994),
primordial perturbations are provided by {\it seeds} of trapped energy density 
produced during
symmetry breaking phase transitions in the early universe. Such symmetry 
breaking is predicted by
Grand Unified Theories (GUT's) to occur at early times as the universe cools and 
expands. The
geometry of the produced seeds, known as {\it topological defects} is determined 
by the topology of
the vaccuum manifold of the physically realized GUT. Thus the defects may be 
pointlike (monopoles),
linelike (cosmic strings), planar (domain walls) or collapsing pointlike 
(textures).
 
The cosmic string theory \markcite{v81}(Vilenkin 1981) for structure formation is the oldest 
and (together with
textures \markcite{t89}(Turok 1989)) best studied theory of the topological defect class. By 
fixing its single free
parameter $G\mu$ ($\mu$ is the {\it effective} mass per unit length of the 
wiggly string and $G$ is
Newtons constant) to a value consistent with microphysical requirements coming 
from GUT's, the
theory may automatically account for large scale filaments and sheets 
\markcite{v86,s87,pb90,vv91,v92,hm93}(Vachaspati 1986;
Stebbins {\it et. al.} 1987; Perivolaropoulos, Brandenberger \& Stebbins 1990; 
Vachaspati \& Vilenkin 1991; Vollick 1992; Hara \& Miyoshi 1993), galaxy
formation at epochs $z\sim 2-3$ \markcite{bk87}(Brandenberger {\it et. al.} 1987) and galactic 
magnetic fields
\markcite{v92b}(Vachaspati 1992b). It
can also provide large scale peculiar velocities \markcite{v92a,pv94}(Vachaspati 1992a; 
Perivolaropoulos \& Vachaspati 1994) and is consistent with the amplitude, 
spectral index
\markcite{bbs88,bs92,p93a}(Bouchet, Bennett \& Stebbins 1988; Bennett, Stebbins \& Bouchet 1992;  
Perivolaropoulos 1993a) and the statistics \markcite{g90,p93b,mp94,cf94,l94,m95b}(Gott {\it et. al.} 1990; 
Perivolaropoulos 1993b; Moessner,
Perivolaropoulos \& Brandenberger 1994;  Coulson {\it et. al.} 1994; Luo 1994;
Magueijo 1995b)
 of the cosmic microwave background (CMB) anisotropies measured by the COBE 
collaboration
\markcite{s92,w92}(Smoot {\it et. al.} 1992; Wright {\it et. al.} 1992) on large angular scales 
($\theta\sim
10^\circ$). Other planned CMB experiments \markcite{m97,cs97,ss95}(see e.g. MAP 1997, COBRAS/SAMBA 1997, see also the review by Scott et. al. 1995) of equally high quality but on smaller angular scales are expected to provide a wealth of information within the next few years.
 
The CMB observations provide a valuable direct probe for identifying signatures 
of cosmic strings.
The main mechanism by which strings can produce CMB fluctuations on angular
scales larger than 1-2 degrees 
has been well studied both analytically \markcite{bt86,s88,vs90,p93a,p93b,mp94}(Brandenberger \& Turok 1986; 
Stebbins 1988; Veeraraghavan \& Stebbins 1990; Perivolaropoulos 1993a; 
Perivolaropoulos 1993b;
Moessner {\it et. al.} 1994) and
using numerical simulations \markcite{bbs88,bs92}(Bouchet {\it et.al.} 1988; 
Bennett {\it et. al.} 1992) and is known as the {\it
Kaiser-Stebbins effect} \markcite{ks84,g85}(Kaiser \&  Stebbins 1984; Gott 1985). According to this
effect, moving long strings present between the time of recombination $t_{rec}$ 
and the present time
$t_0$, produce step-like temperature discontinuities between photons that reach 
the observer through
opposite sides of the string. These discontinuities are due to the peculiar 
nature of the spacetime
around a long string which even though is {\it locally} flat, {\it globally} has 
the geometry of
a cone with deficit angle $8\pi G\mu$. The magnitude of the discontinuity is 
proportional to the
deficit angle, to the string velocity $v_s$ and depends on the relative 
orientation between the unit
vector along the string ${\hat s}$ and the unit photon wave-vector ${\hat k}$. 
It is given by
\markcite{s88}(Stebbins 1988)
\begin{equation} 
{{\delta T}\over T}=\pm 4\pi G\mu v_s \gamma_s {\hat k} \cdot ({\hat v_s}\times 
{\hat s})
\end{equation}
where $\gamma_s$ is the relativistic Lorentz factor and the sign changes when 
the string is
crossed. The angular scale over which this discontinuity persists is given by 
the radius of
curvature of the string which according to simulations \markcite{bb88,as90,as93}(Bennett \& Bouchet 1988; Allen \& 
Shellard 1990;  Albrecht \& Stebbins 1993) is approximately equal
to the horizon scale. 
The growth of the horizon from $t_{rec}$ to $t_0$ results in a superposition
of a large number of step-like temperature seeds of all sizes starting from
about $2^\circ$ (the angular size of the horizon at $t_{rec}$) to about 
$180^\circ$ (the present horizon scale). By the central limit theorem this
large number of superposed seeds results in a pattern of fluctuations that
obeys Gaussian statistics. Thus the probability distribution for the 
temperature of each pixel of a CMB map with resolution larger than about
$1^0 - 2^0$ is a Gaussian  \markcite{ac96,cf94,p93b,p93c}(Allen et. al. 1996, Coulson et. al. 1994, Perivolaropoulos 1993b; Perivolaropoulos 1993c). 
It has therefore been considered to be impossible
to distinguish structure formation models based on cosmic strings from 
corresponding models based on inflation, using CMB maps with resolution 
angle larger than $1^0 -2^0$ \markcite{fm97}(Ferreira \& Magueijo 1997) . Theoretical studies have therefore focused
on identifying the statistical signatures of cosmic strings on angular scales
less than $1^0$ \markcite{t96,cf94}(Turok 1996, Coulson et. al. 1994) where the number of superposed seeds is smaller and therefore
the non-Gaussian character of fluctuations is expected to be stronger 
\footnote{The non-Gaussian features for texture maps are stronger than thosed
of cosmic strings mainly because of the generically smaller number of
textures per horizon volume \markcite{g96,m95a}(Gangui 1996; Magueijo 1995a).}

These efforts however  have been faced with the complicated and model
dependent physical processes occuring on small angular scales. Such 
effects include isolated foreground point sources, recombination physics,
string properties on small scales (kinks, loops etc) which require
detailed simulations of both the string network and the cosmic background,
in order to be properly taken into account.  Even though preliminary
efforts for such detailed simulations are in progress \markcite{ac96}(Allen et. al. 1996), it has become clear
that it will take some time before theory and experiments on angular
scales less than a few arcmin reach accuracy levels leading to detectable
non-Gaussian string signatures.

 An alternative approach to the problem is instead of focusing on small
scales where the number of superposed seeds is small, to focus on larger
angular scales where despite the large number of superposed seeds there 
is also coherence of induced fluctuations on large angular scales. Fluctuations
on these scales may be viewed as a superposition of a Gaussian scale 
invariant background coming mainly from small scale seeds plus a small 
number of step-like discontinuities which are coherent and persist on angular 
scales larger than $100^0$. These are produced by long strings present in
our present horizon. {\it Our goal is to find a statistic optimized to detect 
this large scale coherence and use it to find the minimum amplitude of a step 
function that can be detected at the $1\sigma$ level relatively to a given scale 
invariant Gaussian background.}
Such a statistic is equally effective on {\it any} angular resolution scale and 
its effectiveness is only diminished as the number of pixels of the CMB map is 
reduced or the noise is increased. 
The statistical variable we focus on, in what follows is the Sample Mean 
Difference (SMD) between large neighbouring sectors of a CMB map. These sectors 
are separated by a random straight line in two dimensional maps or by a random 
point  in one dimensional maps. The union of the two sectors gives back the 
complete map. 
We show that the statistics of the SMD variable are much more sensitive in 
detecting the presence of a coherent step-like seed than conventional statistics 
like the skewness or the kurtosis.

The structure of this paper is the following: 
In the next section titled 'Sample Mean Difference' we study analytically the 
statistics of the SMD variable and show that its average value is a sensitive 
quantity in detecting the presence of a randomly positioned step-function on top 
of a gaussian map. We then compare with the sensitivity of the statistics {\it 
skewness} and {\it kurtosis}. We find that the sensitivity of the SMD statistics 
is significantly superior to that of skewness and kurtosis in detecting the step 
function. These analytical results are shown for the case of one-dimensional 
maps but the extension to the case of two dimensional maps is straightforward.

In the third section titled 'Monte Carlo Simulations' we perform Monte Carlo 
simulations of Gaussian maps with flat and scale invariant power spectra, with 
and without step-like discontinuities in one and two dimensions. We also include 
noise with signal to noise ration  ${s\over n}=1.0$. Applying the statistics 
skewness, kurtosis and average of SMD on these maps we verify the results of 
section 2 and find the minimum step-function amplitude that is detectable by the 
average SMD statistic. Finally is section 4 we conclude, summarise and discuss 
the prospect of applying the mean of SMD statistic to presently available CMB 
maps including the COBE results. That analysis is currently in progress 
\markcite{ps97}(Athanasiou,Perivolaropoulos \& Simatos 1997).

\section{\bf Sample Mean Difference}

Consider an one dimensional array of $n$ pixel variables $x_n$. Let these 
variables be initially distributed according to a standardised Gaussian 
probability distribution. Consider now a step-function of amplitude $2\alpha$  
superposed so that the discontinuity is between pixels $i_0$ and $i_0 + 1$ (Fig. 
1). The new probability distribution for a random pixel variable $x$ is 
\be
P(x) = {f \over {\sqrt{2\pi}}} e^{-{{(x-\alpha)^2} \over 2}} + 
{{1-f} \over {\sqrt{2\pi}}} e^{-{{(x+\alpha)^2} \over 2}}
\ee
where $f = {{i_0} \over n}$. We are looking for a statistic that will optimally 
distinguish between a Gaussian array with a superposed step-function and a 
Gaussian array without one. The obvious statistics to try first are the moments 
of the distribution (2) with $\alpha = 0$ and $\alpha \neq 0$.

The moment generating function corresponding to (2) is:
\be
M(t) = f e^{\alpha t + {t^2 \over 2}} +
(1-f) e^{-\alpha t + {t^2 \over 2}}
\ee
The mean $\mu (\alpha, f)$, variance $\sigma^2 (\alpha,f)$, skewness 
$s(\alpha,f)$ and kurtosis $k(\alpha,f)$ can be obtained in a straightforward 
way by proper differentiation of $M(t)$ as follows:
\ba
\mu (\alpha, f)&\equiv &<X> = \alpha f - \alpha (1-f) \\
\sigma^2 (\alpha,f) &\equiv & < (x - \mu )^2 > = 
1 + 4 \alpha^2 f (1 - f) \\
s(\alpha,f) &\equiv & {{< (x - \mu )^3 >}\over \sigma^3} =
{{8 \alpha^3 f (1-3 f + 2 f^2)} \over {(1 + 4 \alpha^2 f (1 - f))^{3/2}}} 
\ea
\begin{eqnarray*}
k(\alpha,f) &\equiv &  {{< (x - \mu )^4 >}\over \sigma^4}\\ &=&
{{3 + 8 \alpha^2 f (3 + 2\alpha^2 -3f^2 - 8 \alpha^2 f + 12 \alpha^2 f^2 
-6\alpha^2 f^3)} \over {(1 + 4 \alpha^2 f (1 - f))^4}}
\end{eqnarray*}

For $\alpha = 0$ we obtain the Gaussian values for the skewness and the kurtosis 
$s(0,f) = 0$ , $k(0,f) = 3$ as expected. For $\alpha \neq 0$ the moments deviate 
from the Gaussian values. In order to find the minimum value of $\alpha$ for 
which the moments can distinguish between a Gaussian pattern and a Gaussian+Step 
pattern we must compare the deviation of moments from their Gaussian values with 
the standard deviation of the sample moments.
The mean values of the skewness and the kurtosis are easily obtained by 
integrating with respect to $f$ from 0 to 1 i.e. assuming that it is equally 
probable for the step-function to be superposed at any point of the lattice.
\ba
{\bar s}(\alpha) & = &<s(\alpha, f)> = \int_0^1 df s(\alpha,f) = 0 \\
{\bar k}(\alpha) & = &<k(\alpha, f)> = \int_0^1 df k(\alpha,f)
\ea
These values are to be compared with the standard deviations of the moments, 
obtained as follows:
The variance of the skewness over several n-pixel array realizations with fixed 
$f$ and $\alpha$ is
\be
\Delta s^2 (\alpha,f) = <({\hat s} - s)^2 >
\ee
where ${\hat s} \equiv {{s_1 + ... + s_n} \over n}$ is the sample skewness from 
a given pixel array realization, $s$ is the actual skewness and $s_i \equiv 
{{(x_i - \mu)^3} \over \sigma^3}$. Now
\be
<{\hat s}> = {{n < s_1>} \over n} = <s_1> = s
\ee
Also
\be
<{\hat s}^2> = {1\over n} <s_j^2> + (1-{1\over n}) <s_j>^2
\ee
where $j$ {\it any} pixel number ($j \in [1,n]$).
Thus
\be
\Delta s^2 (\alpha,f) = {1\over n} (<s_j^2> - <s_j>^2) = 
{1\over n}{1 \over \sigma^6} <(x_j - \mu)^6>
\ee
Similarly for the variance of the sample kurtosis we have
\be
\Delta k^2 (\alpha,f) = {1\over n} (<k_j^2> - <k_j>^2)  
\ee
with $k_j=
{1 \over \sigma^4}  (x_j - \mu)^4
$
and  
$  
<k_j^2> = {1 \over \sigma^8} <(x_j-\mu)^8>
$.
It is straightforward to obtain all the above moments by differentiating
the generating functional and using
\be
< x_j^n > ={{d^n M} \over {d t^n}} \vert_{t=0} 
\ee
Now the minimum value $\alpha_{min}$ of $\alpha$ detectable at $1\sigma$
level is obtained from the equations
\ba
\int_0^1 df [s(\alpha_{min},f) - \Delta s(\alpha_{min},f)] &=& 0 \\
\int_0^1 df [(k(\alpha_{min},f)-3) - \Delta k(\alpha_{min},f)] &=& 0  
\ea
Since (from eq. (7)) ${\bar s} (\alpha) = 0$ which is equal to the 
Gaussian value, the skewness can only be used to detect a step function
by comparing the standard deviation $\Delta{\bar s}$ for $\alpha = 0$ and
$\alpha \neq 0$. By demanding $\Delta {\bar s} (\alpha_{min}) \leq 2 
\Delta {\bar s} (\alpha = 0)$ we obtain $\alpha_{min} \leq 2.5$. This
result is independent of the number of pixels $n$. For the kurtosis
we obtain from eqs. (13, 16) $\alpha_{min} \simeq 4 $ for $n = 10^3$ while
for $\alpha_{min} =0.5$, $n\simeq 10^6$ is required.

Using the alternative test i.e. demanding $\Delta {\bar k} (\alpha_{min}) 
\geq 2 \Delta {\bar k} (\alpha = 0) $ we obtain $\alpha \geq 2$ and this
result is independent of the number of pixels n as in the case of skewness.
Thus for the usual pixel maps where n is up to $O(1000)$ the kurtosis is
not able to detect a step function with $\alpha \leq 2$ at the $1\sigma$
level. As in all cases discussed in this paper $\alpha$ is measured
in units of standard deviation (rms) of the underlying Gaussian map.
This result remains unchanged for other statistical variables defined by {\it local} linear combinations of pixels (e.g. differences of neighbouring pixel variables \markcite{mp94,cf94}(Moessner et. al. 1994, Coulson et. al. 1994)) since the effect of a single discontinuity remains negligible if the long range coherence is not taken into account.

For CMB temperature maps with $({{\delta T} \over T})_{rms} \simeq 
2 \times 10^{-5}$ the detectable value of $G\mu$ is
\be
\alpha \equiv 4\pi G \mu (v_s \gamma_s)\cos \theta > 4 \times 10^{-5} 
\Rightarrow
G \mu (v_s \gamma_s)\cos \theta \geq 4 \times 10^{-6}
\ee
where $\theta$ is an angle obtained from the relative orientation of the string 
with respect to the observer.
According to simulations $<v_s \gamma_s >_{rms} \simeq 0.2 $ and for
$G\mu < 2 \times 10^{-5}$ the detection of the Kaiser-Stebbins effect 
using statistics based on skewness and kurtosis is not possible. This excluded range 
however includes all the cosmologically interesting 
values of $G\mu$.

It is therefore important to look for alternative statistical
variables that are more sensitive in detecting the presence
of coherent discontinuities superposed on Gaussian maps. As we will
show, the Sample Mean Difference (SMD) is such a statistical variable. 

Consider a pixel array (Fig. 1) of $n$ pixel Gaussian random variables
$X_j$ with a step function covering the whole array, superposed such 
that the discontinuity is located just after pixel $i_0$. To every pixel
$k$ of the array we may associate the random variable $Y_k$ defined as
the difference between the mean value of the pixels 1 through $k$ minus the 
mean value of the pixels $k+1$ through $n$.
It is straightforward to show that
\ba
Y_k &=& \Delta {\bar X}_k + 2 \alpha {{n-i_0} \over {n-k}} \hspace{1cm}
k\in [1,i_0] \\
Y_k  &=& \Delta {\bar X}_k + 2 \alpha {i_0 \over k} \hspace{1cm}
k\in [i_0 ,n-1] 
\ea
where $\Delta {\bar X}_k ={1\over k} \sum_{j=1}^k X_j - {1 \over {n-k}}
\sum_{j=k+1}^n X_j$. Thus we have constructed a new array $Y_k$, $(k=1,...,n-1)$ 
from the sample mean differences (SMD) of the original array. We will focus on 
the average value $Z$ of the SMD defined as:
\be
Z={1\over {n-1}} \sum_{k=1}^{n-1} Y_k
\ee
Using eqs. (19,20) we obtain
\be
Z={1\over {n-1}} [\sum_{k=1}^{n-1} \Delta{\bar X}_k + 2\alpha 
(\sum_{k=1}^{i_0} {{1-i_0 / n} \over {1-k/n}} + 
\sum_{k=i_0 + 1}^{n-1} {{i_0/n} \over {k/n}}) ]
\ee
With the definitions $f\equiv {i_0 / n}$ and $\xi \equiv {k/n}$ and the
assumption $n>>1$ we obtain:
\be
Z = {1 \over {n-1}} \sum_{k=1}^{n-1} \Delta {\bar X}_k - 2\alpha 
[(1-f) \ln(1-f) + f \ln f]
\ee
Thus the mean of $Z$ over many realizations of the array is
\be
<Z> = {1 \over {n-1}} \sum_{k=1}^{n-1} <\Delta {\bar X}_k>- 4\alpha 
[\int_0^1 df \hspace{1mm} f \ln f] = \alpha
\ee
The variance of $Z$ is due both to the underlying Gaussian map and to the 
variation of $f=i_0 /n $ (assuming $\alpha$ fixed). The variance due to the  
gaussian background is
\be
\sigma_{1,Z} = {1\over {(n-1)^2}} \sum_{k=1}^{n-1} ({1\over k} + 
{1\over {n-k}}) \simeq \epsilon \int_\epsilon^{1-\epsilon} {{d\xi} \over {\xi 
(1-\xi)}}
\ee
where $\epsilon = O({1\over n})$, $\xi = k/n$, $n>>1$ and we have used the fact 
that the variance of the sample mean of a standardized Gaussian population with 
size $j$ is  $1\over j$. Now from eq. (24) we obtain
\be
\sigma_{1,Z}^2 \simeq -\epsilon \ln \epsilon^2 \simeq {{2 \ln n}\over n}
\ee
The variance of the $f$-dependent part of Z is
\be
\sigma_{2,Z} = <Z_2^2> - <Z_2>^2
\ee
where $Z_2 \equiv -2\alpha [(1-f) \ln(1-f) + f \ln f] $. From eq. (23) we have 
$<Z_2> = \alpha$ and $<Z_2^2>$ is easily obtained as
\be
<Z_2^2> = \int_0^1 df \hspace{1mm} Z_2^2 (f) \simeq {4\over 3} \alpha
\ee
Thus
\be
\sigma_Z^2 \equiv \sigma_{1,Z}^2 + \sigma_{2,Z}^2 = {{2\ln n} \over n} + {1\over 
3}\alpha^2
\ee
In order to be able to distinguish between a Gaussian+Step map and a purely 
Gaussian one, at the $1\sigma$ level we demand that
\be
<Z>_{\alpha \neq 0} - <Z>_{\alpha = 0} \geq \sigma_Z
\ee
This implies that the minimum value of $\alpha$, $\alpha_{min}$ that can be 
detected using this test is
\be
\alpha_{min} = ({{3\ln n}\over n})^{1/2}
\ee
and for $n=O(10^3)$ we obtain $\alpha_{min} \simeq 0.2$ which is about an order 
of magnitude improvement over the corresponding sensitivity of tests based on 
the moments skewness and kurtosis. The reason for this significant improvement 
is the fact that the SMD statistical variable picks up the {\it coherence} 
properties introduced by the step function on the Gaussian map. The moments on 
the other hand pick up only local properties of the pixels and do not amplify 
the long range coherence of the step-like discontinuity.

Our analysis so far has assumed that the Gaussian variables $X_j$ are 
independent and that the only correlation is introduced by the superposed 
step-function. In a realistic setup however the underlying Gaussian map will be 
scale invariant and thus there will be correlations among the pixels. These 
correlations will also be affected by the instrument noise. In addition, our 
analysis has been limited so far to one dimensional maps while most CMB 
experiments are now obtaining two-dimensional maps. In order to take all these 
effects into account we need to apply the statistics of the SMD variable onto 
maps constructed by Monte Carlo simulations. This is the focus of the following 
section. 

\section{\bf Monte-Carlo Simulations}
We start by constructing an array of n Gaussian random variables $X_j$, 
$j=1,...,n$ with a power spectrum $P(k)=k^{-m}$. Thus the values $X_j$ associated 
with the pixel $j$ is obtained as the Fourier transform of a function $g(k)$ 
($k=1,...,n$) with the following properties:
\begin{itemize}
\item
For each $k$, the amplitude $|g(k)|$ is an independent random variable with 0 
mean and variance $P(k) = 1/k^m$.
\item
The phase $\theta_k$ of each Fourier component $g(k)$ is an independednt random 
variable in the range $[0,2\pi]$ with uniform probability distribution 
$P(\theta_k) = {1\over {2 \pi}}$.
\item
The Fourier components are related by complex conjugation relations neeeded to 
give a {\it real} variable $X_j$.
\end{itemize}
The discrete Fourier transform definition used is
\be
X_j = {1\over \sqrt{n}} \sum_{k=1}^n g(k) e^{2\pi i (k-1)(j-1)/n}
\ee
and the numerical programming was implemented using {\it Mathematica}
\markcite{w91}(Wolfram 1991).
In order to have real $X_j$, the conditions 
\ba
\rm{Im} g(0) &=& 0 \\
g(k+2) &=& g^* (n-k) \hspace{1cm} k=0,...,n-2 \\
\rm{Im} g({n\over 2} +1) & = & 0 
\ea
must be satisfied. The array $X_j$ obtained in the way described above is then 
standardized to a new array $X_j^s$, with 
\be
X_j^s \equiv {{(X_j - \mu)}\over {\sigma^2}}
\ee
where $\mu$ and $\sigma^2$ are the sample mean and sample variance for the 
realization of the array $X_j$. A new array $X_j^\prime$ is then constructed by 
superposing to the array $X_j^s$ a step function of amplitude $2\alpha$ with 
discontinuity at a random point $i_0$. The array $X_j^\prime$ is thus obtained 
as
\be
X_j^\prime = X_j^s + \alpha {{j-i_0} \over {|j-i_0|}}, \hspace{1cm} j=1,...,n
\ee
Next we apply the statistics discussed in the previous section to several 
realizations of the arrays $X_j^s$ and $X^\prime$ in an effort to find the most 
sensitive statistic that can distinguish among them. Our goal is to also find 
the minimum value of $\alpha$ that can be distinguished by that statistic at the 
$1\sigma$ level, thus testing the analytical results of the previous section.

We have used a lattice with 2000 pixels and a scale invariant power spectrum 
which for one-dimesional data is $P(k)=k^{-1}$.
In Table 1 we show the results for the skewness, the kurtosis and the average 
SMD for the $X_j$ arrays, with $\alpha=$0, 0.25, 0.50 and 1.0. The SMD average 
was obtained as in section 2 by first constructing the array of sample mean 
differences and then obtaining its average value, predicted to be equal to 
$\alpha$ by the analytical study of section 2.

These statistics were applied to 50 random realizations of the array $X_j^s$. 
The mean values of the statistics considered with their $1\sigma$ standard 
deviations obtained over these 50 realizations are shown in the following Table 
1.
\newpage

{\bf Table 1}: A comparison of the effectiveness of the statistics considered, in 
detecting the presence of a coherent step discontinuity with amplitude $2\alpha$ 
relative to the standard deviation of the underlying Gaussian map. No noise was 
included in these simulations and the full map was used in obtaining the SMD 
average.
\vskip 0.1cm
\begin{tabular}{|c|c|c|c|}\hline
{\bf $\alpha$ }&{\bf  Skewness }&{\bf  Kurtosis }&{\bf SMD Average } \\ \hline
0.00 &$0.01 \pm 0.11$     &$2.97 \pm 0.19$     &$0.02 \pm 0.31$\\ \hline
0.25 &$0.01 \pm 0.11$     &$2.95 \pm 0.20$     &$0.25 \pm 0.33$ \\ \hline
0.50 &$0.02 \pm 0.11$     &$2.88 \pm 0.21$     &$0.48 \pm 0.38$ \\ \hline
1.00 &$0.03 \pm 0.20$     &$2.82 \pm 0.32$     &$0.98 \pm 0.48$ \\ \hline
\end{tabular}
\vspace{3mm}

The analytical prediction of section 2 for the SMD average value $\alpha$ is in 
good agreement with the results of the Monte Carlo simulations. The standard 
deviation of this result is not in such a good agreemnent with the analytical 
prediction because the assumption of complete independence among pixels made by 
the analytical treatment is not realized in the Monte Carlo simulations where a 
scale invariant spectrum was considered and thus there was a non-trivial 
correlation among the pixels of the arrays.

The effects of adding uncorrelated Gaussian noise with signal to noise ratio 
${s\over n} = 1$ are shown in Table 2. This table was constructed by adding an 
uncorrelated Gaussian signal of unit variance to the standardized forms of the 
arrays $X_j^s$ and $X_j^\prime$ and then repeating the statistics of Table 1.

From the results of Table 2 it becomes clear that the effects of noise do not 
affect significantly the sensitivity of the SMD average in detecting the 
presence of the coherent step. 
\newpage

{\bf Table 2}: Similar to Table 1 but including noise in the Monte Carlo 
simulations with signal to noise ratio ${s\over n}=1.0$. The full map was used 
in obtaining the SMD average.
\vskip 0.1cm
\begin{tabular}{|c|c|c|c|}\hline
{\bf $\alpha$ }&{\bf  Skewness }&{\bf  Kurtosis }&{\bf SMD Average } \\ \hline
0.00 &$0.01 \pm 0.10$     &$2.97 \pm 0.15$     &$0.04 \pm 0.31$\\ \hline
0.25 &$0.01 \pm 0.09$     &$2.98 \pm 0.13$     &$0.21 \pm 0.34$ \\ \hline
0.50 &$0.02 \pm 0.08$     &$2.98 \pm 0.14$     &$0.42 \pm 0.39$ \\ \hline
1.00 &$0.01 \pm 0.10$     &$2.92 \pm 0.14$     &$0.97 \pm 0.53$ \\ \hline
\end{tabular}
\vspace{3mm}

A simple way to further improve the sensitivity 
of the SMD statistical variable is to ignore a number $l$ of boundary pixels of 
the SMD array, thus constructing its average using the Sample Mean Differences 
of pixels $l+1,...,n-l$. From eq (24), the variance of the SMD  for these pixels 
is significantly lower than the corresponding variance of the $2l$ pixels close 
to the boundaries. In addition, if the step is located within the central $n-2l$ 
pixels the SMD average may be shown to be larger than $\alpha$ thus further 
amplifying the step signature. For $l=150$ the variance of the SMD average {\it 
is reduced} by about $20$\% (Table 3) while the SMD average is {\it increased} by 
about $20$\% thus allowing the detection of steps as low as $\alpha = 0.25$ at 
the $1\sigma$ level. The price to pay for this sensitivity improvement is the 
reduction of the effective pixel area where the search for steps is made.

We have also used the SMD statistical variable for non-scale invariant power 
spectra and found that it works better for $P(k) = k^{-m}$ with $0 \leq m <1$ 
than for $m>1$. This is to be expected because large values of $m$ imply larger 
correlations among pixels which in turn leads to a smaller number of effectively 
independent pixels and thus a larger value for the variance of the SMD average. 

\newpage
{\bf Table 3}: Similar to Table 1 but the SMD average was obtained after 
ignoring 150 pixels on each boundary of the Monte Carlo maps. The 
discontinuities were also excluded from these 300 pixels. This significantly 
improved the sensitivity of the SMD test.
\vskip 0.1cm
\begin{tabular}{|c|c|c|c|}\hline
{\bf $\alpha$ }&{\bf  Skewness }&{\bf  Kurtosis }&{\bf SMD Average } \\ \hline
0.00 &$0.01 \pm 0.10$     &$2.96 \pm 0.15$     &$0.01 \pm 0.24$\\ \hline
0.25 &$0.01 \pm 0.09$     &$2.95 \pm 0.15$     &$0.28 \pm 0.26$ \\ \hline
0.50 &$0.02 \pm 0.14$     &$2.94 \pm 0.18$     &$0.63 \pm 0.29$ \\ \hline
1.00 &$0.03 \pm 0.30$     &$2.78 \pm 0.30$     &$1.21 \pm 0.46$ \\ \hline
\end{tabular}
\vspace{3mm} 

It is straightforward to generalize the one dimensional Monte Carlo simulations 
to two dimensions. In that case we use the two-dimensional discrete Fourier 
transform as an approximation to an expansion to spherical harmonics. This 
approximation is good for small area maps of the celestial sphere. We used the 
following definition of the two dimensional discrete Fourier transform.
\be
X(i,j) = {1 \over n} \sum_{k_1 = 1}^n \sum_{k_2 = 1}^n g(k_1,k_2) e^{2\pi 
i[(i-1) (k_1 -1) +(j-1) (k_2 - 1)]/n}
\ee
refering to a $n \times n$ square lattice. In order to construct the background 
of scale invariant Gaussian fluctuations we used $g(k_1,k_2)$ as a complex 
random variable. For scale invariance, the amplitude of $g(k_1,k_2)$ was 
obtained from a Gaussian probability distribution with 0 mean and variance
\be
\sigma^2 (k_1,k_2) = P(k_1,k_2) = {1\over {k_1^2 + k_2^2}}
\ee
The corresponding phase $\theta_{k_1,k_2}$ for the $(k_1,k_2)$ mode was also 
determined randomly from a uniform probability distribution 
$P(\theta_{k_1,k_2})={1\over {2\pi}}$ in order to secure Gaussianity for the map 
$X(i,j)$. To secure that the Fourier transformed map $X(i,j)$ consists of real 
numbers, the following constraints were imposed on the spectrum $g(k_1,k_2)$
\begin{eqnarray*}
{Im} g(1,1) &=& {Im} g({n\over 2} + 1, 1)=\\
={Im} g(1,{n\over 2} + 1) &=&  { 
Im} g({n\over 2} + 1,{n\over 2} + 1)=0 \\
g(n-i,1) &=& g^* (i+2,1), \hspace{3.0cm} i=0,...,{n\over 2} -2 \\
g(1,n-i) &=& g^* (1,i+2), \hspace{3.0cm} i=0,...,{n\over 2} -2 \\
g(i,j) &=& g^* (n-i+2,n-j+2), \hspace{0.8cm} i=2,...,{n\over 2} +1,\\
&  &\hspace{5cm} j=2,...,n  
\end{eqnarray*}

The corresponding map with a superposed coherent step discontinuity was obtained 
from the standardized Gaussian map $X^s (i,j)$ as
\be
X^\prime (i,j) = X^s(i,j) + \alpha {{j-a \hspace{1mm} i - b} \over
{|j-a \hspace{1mm} i - b|}}
\ee
where 
\ba
a&=&{{y_2-y_1} \over {x_2 - x_1}}\\
b &=& y_1 - a \hspace{1mm} x_1
\ea
i.e. the line of step discontinuity $j=a \hspace{1mm} i +b$ is determined by the 
two random points $(x_1,y_1)$ and $(x_2,y_2)$ of the map $X(i,j)$. The skewness 
and kurtosis of the two maps are obtained in the usual way. For example for the 
standardized Gaussian map $X^s (i,j)$ we have
\ba
s &=& {1\over n^2} \sum_{i,j}^n X^s (i,j)^3 \\ 
k &=& {1\over n^2} \sum_{i,j}^n X^s (i,j)^4
\ea
The SMD statistical variables is obtained by considering a set of random 
straight lines bisecting the map and for each line taking the difference of the 
sample means from the two parts of the map. For example consider a line defined 
by the random points $(x_1,y_1)$ and $(x_2,y_2)$ of the map. The line equation 
is $j=a \hspace{1mm} i + b$ with $a$, $b$ obtained from eqs. (40) and (41). The 
SMD obtained from this line is
\be
\rm{SMD} = {S_1 \over n_1} - {S_2 \over n_2}
\ee
where
\ba
S_1 &=& \sum_{i=1}^n \sum_{j=Max[(a \hspace{1mm} i + b),1]}^n X^s(i,j) \\     
S_2 &=& \sum_{i=1}^n \sum_{j=1}^{Min[(a \hspace{1mm} i + b),n]} 
X^s(i,j)    
\ea 
and  $n_1$, $n_2$ are the corresponding numbers of terms in the sums. For a 
Step+Gaussian map, the discontinuity the index $^s$ get replaced by $^\prime$. 

The average and variance of the SMD  
is obtained by averaging over a large number 
of random test lines $(a,b)$ and a large number of map realizations. The results 
of the application of the three statistics (skewness, kurtosis and SMD average) 
on $30 \times 30$ scale invariant Gaussian maps for various values of step 
amplitudes $\alpha$ are shown in Table 4. Uncorrelated Gaussian noise with 
signal to noise ratio ${s\over n} = 2.0$ was also included. The
random points defining the test lines were excluded 
from the outermost three rows and columns of the maps thus reducing somewhat the 
variance of the SMD average.      

{\bf Table 4}: A comparison of the effectiveness of the statistics considered in 
two dimensional maps. A signal to noise ratio of ${s\over n}=2.0$ was assumed in 
these maps. Points defining the line discontinuities were excluded from the three outermost rows and columns of the maps.
\vskip 0.1cm
\begin{tabular}{|c|c|c|c|}\hline
{\bf $\alpha$ }&{\bf  Skewness }&{\bf  Kurtosis }&{\bf SMD Average }\\ \hline
0.00 &$0.04 \pm 0.13$     &$3.00 \pm 0.20$     &$0.01 \pm 0.03$\\ \hline
0.25 &$0.02 \pm 0.08$     &$2.97 \pm 0.13$     &$0.14 \pm 0.09$ \\ \hline
0.50 &$0.05 \pm 0.14$     &$2.91 \pm 0.24$     &$0.34 \pm 0.19$ \\ \hline
1.00 &$0.02 \pm 0.24$     &$2.95 \pm 0.30$     &$0.56 \pm 0.31$ \\ \hline
\end{tabular}
\vspace{3mm}

The results of Table 4 are in qualitative agreement with those of Tables 1-3 and 
with the analytical results valid for the one dimensional maps. 
Clearly the details of the one dimensional analysis are not valid in the two 
dimensional case and so the agreement can not be quantitative.
The results still indicate however that the SMD statistic is significantly more 
sensitive compared to conventional statistics for the detection of coherent 
discontinuities on CMB maps. This statistic can detect coherent discontinuities 
with minimum amplitude $\alpha_{min} \simeq 0.5$ at the $1\sigma$ to $2\sigma$ level where 
$\alpha$ is the amplitude relative to the standard deviation of the underlying 
scale invariant Gaussian map.

\section{\bf Conclusion-Outlook}

It is straightforward to apply the statistic analyzed in this paper to realistic 
data of ongoing experiments. Consider for example a $n \times n$ pixel sector of 
the COBE map including $n^2$ pixels. Let also 
$({{\delta T} \over T})_{rms}$ be the $rms$ temperature fluctuations of the 
sector under consideration. The presence of a late long string through this 
sector would have caused a temperature step-discontinuity coherent over the 
whole map, with magnitude $\alpha$ given by eq. (1). If the average of the SMD 
over this sector is found to be very close to 0 (more than $1\sigma$ away from 
the SMD average value for $\alpha = 0.5$), then we may conclude that
\be
{\alpha \over {({{\delta T} \over T})_{rms}}} \leq 
{\alpha \over {({{\delta T} \over T})_{rms}^g}} \leq 0.5
\ee
at the $1\sigma$ confidence level, where $({{\delta T} \over T})_{rms}^g$ is the 
$rms$ value of the purely Gaussian part of the fluctuations which is clearly 
smaller than the total $({{\delta T} \over T})_{rms}$ which includes the step 
discontinuity. Thus
\be
G\mu v_s \gamma_s \cos \theta \leq {1\over {8 \pi}} ({{\delta T} \over T})_{rms}
\ee
where $\theta$ is an angle determined by the orientation of the string with 
respect to the observer. For example for $({{\delta T} \over T})_{rms} = 
10^{-5}$ we obtain $G\mu v_s \gamma_s \cos \theta \leq 4 \times 10^{-7}$
at the $1\sigma$ level. 

Thus using the SMD statistic which is optimized to detect coherent temperature 
discontinuities on top of Gaussian temperature maps we may obtain non-trivial 
upper or even {\it lower} bounds on the values of $G\mu v_s \gamma_s$ which are 
highly robust and independent of the details of the string evolution and the 
resolution of the CMB maps. Application of this statistic on the COBE data is 
currently in progress \markcite{ps97}(Athanasiou, Perivolaropoulos \& Simatos 1997).

\acknowledgments

I wish to thank G. Athanasiou, T. Tomaras and N. Simatos for
interesting discussions and for providing helpful comments after reading the 
paper.
This work was supported by the Greek General Secretariat for Research and 
Technology under grants $\Pi ENE\Delta$  1170/95 and $95 E \Delta 1759$ and by the EEC grants $CHRX-CT93-0340$ and $CHRX-CT94-0621$.

%%%%%%%%%%%%%%%%% Figure 1 %%%%%%%%%%%%%%%%%%
\begin{figure}%%[h]
\begin{center}
\unitlength1cm
                      % see dvips handbook  
\begin{picture}(3,4)
\put(-7.5,8.0){\includegraphics{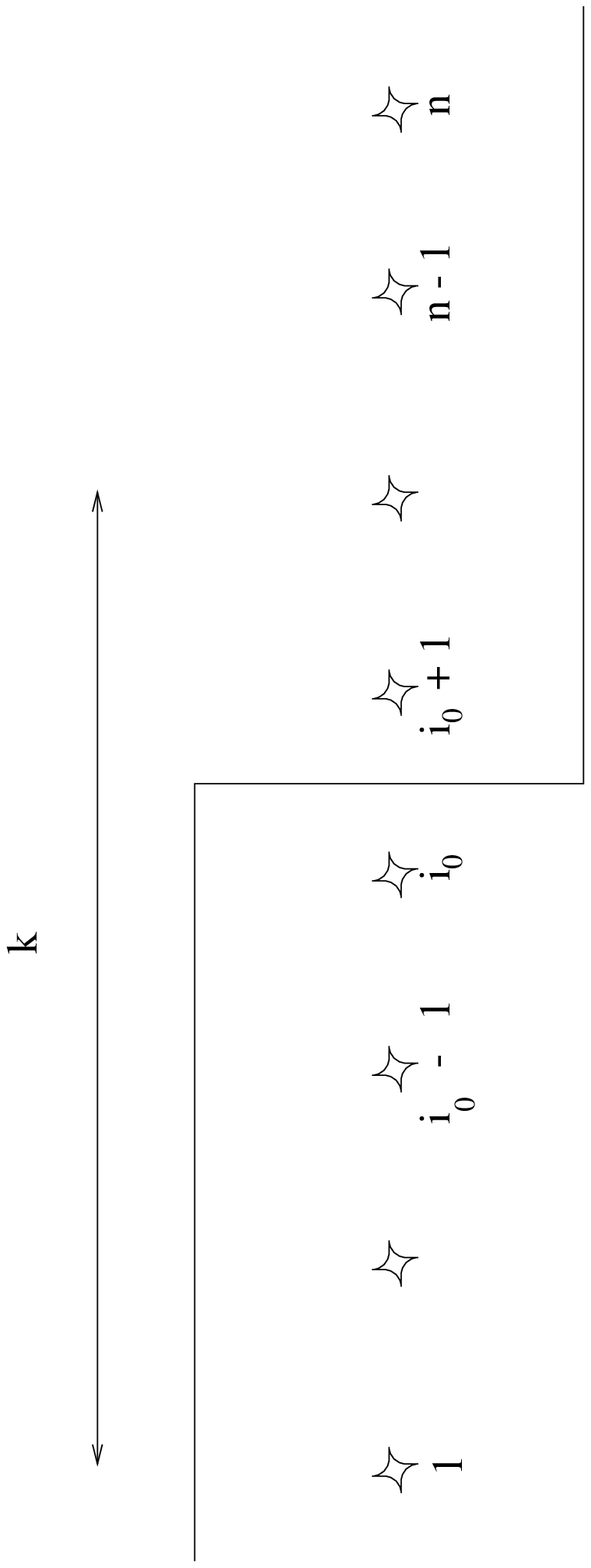}}
\end{picture}
\end{center}
\caption{A large scale coherent step-function discontinuity superposed on a one 
dimensional pixel map.
}
\end{figure}
%%%%%%%%%%%%%%%%%%%%%%%%%%%%%%%%%%%%%%%%%%%%
\vspace{4cm}

%%%%%%%%%%%%%%%%%% Figure 2 %%%%%%%%%%%%%%%%%%
\begin{figure}%%[h]
\begin{center}
\unitlength1cm
                      % see dvips handbook  
\begin{picture}(6,3)
\put(-0.5,-1.8){\includegraphics{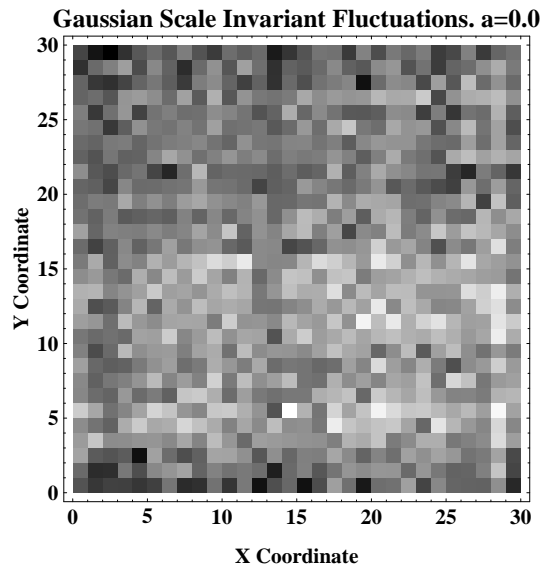}}
\end{picture}
\end{center}
\caption{A standardized two dimensional pixel array of scale invariant Gaussian 
fluctuations. No step function has been superposed.
}
\end{figure}
%%%%%%%%%%%%%%%%%%%%%%%%%%%%%%%%%%%%%%%%%%%%
\vspace{4cm}

%%%%%%%%%%%%%%%%%% Figure 3 %%%%%%%%%%%%%%%%%%
\begin{figure}%%[h]
\begin{center}
\unitlength1cm
                      % see dvips handbook  
\begin{picture}(6,3)
\put(-0.5,-1.8){\includegraphics{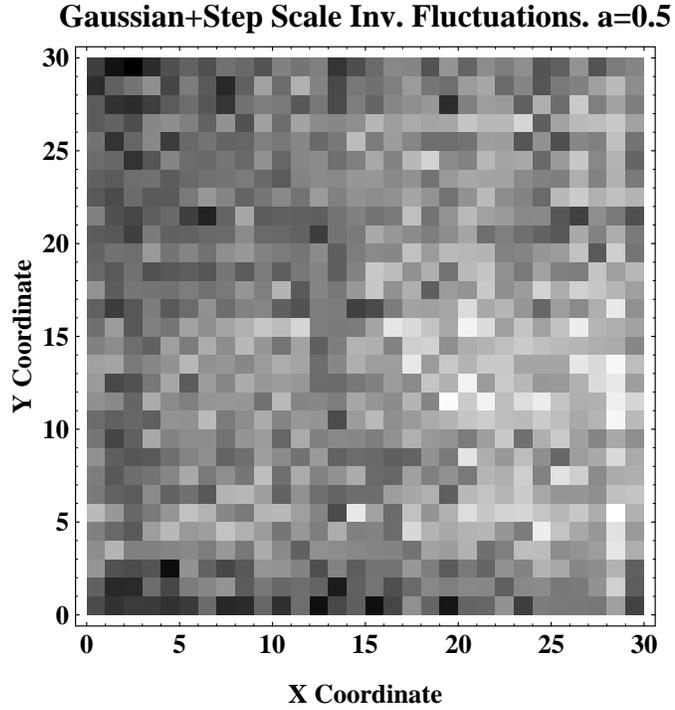}}
\end{picture}
\end{center}
\caption{The two dimesnsional array of Figure 2 with a superposed coherent 
step-discontinuity of amplitude
$\alpha= 0.5$ defined by the random points $(x_1,y_1)=(13.6,18.1)$ and $(x_2,y_2) = (9.4,20.4)$
}
\end{figure}
%%%%%%%%%%%%%%%%%%%%%%%%%%%%%%%%%%%%%%%%%%%%

\end{document}